\documentclass{article}

\def\mytitle#1{\setcounter{equation}{0}
\setcounter{footnote}{0}
\begin{flushleft}\Large\textbf{#1}\end{flushleft}
\vspace{0.25cm}}
\def\myname#1{\leftline{{\large #1}}\vspace{-0.13cm}}
\def\myplace#1#2{\small\begin{flushleft}\textit{#1}\\
\texttt{#2}\end{flushleft}}
\newenvironment{contribution}{\normalsize\noindent}{}

\usepackage{graphicx}
\begin{document}

\mytitle{Evolution of the horizons for dark energy universe}

\frenchspacing \myname{Ritabrata
Biswas,~Nairwita Mazumder,~Subenoy Chakraborty} \frenchspacing

\myplace{Department of Mathematics, Jadavpur University,
Kolkata-32, India.}
{[*schakraborty@math.jdvu.ac.in~,**biswas.ritabrata@gmail.com~,~$\dag$nairwita15@gmail.com]}


\begin{abstract}
Recent observational evidences of accelerating phase of the universe strongly demand that the dominating matter in the universe is in the form of dark energy. In this work, we study the evolution of the apparent and event horizons for various dark energy models and examine their behavior across phantom barrier line.

\end{abstract}

\section{INTRODUCTION}

The prediction of standard cosmology to have at present a phase of deceleration was ruled out in recent past by a series of observations namely the discovery of 16type Ia supernova(SNIa)by Riess et. al. (2004), WMAP(2003) and SDSS(2004). Using the Hubble telescope these observations has provided a distinct scenario of accelerated expansion of the present day universe. Thus a modification of Einstein equations \cite{} becomes essential to incorporate this observational fact. One can either modify the geometry (i.e., the left hand side of Einstein equation ) or the matter itself(i.e., the R.H.S.) if not both. Due to modification of geometry, one can introduce modified gravitytheory namely $f(R)$ gravity, Brane scenario etc while change in the matter part indicates inclusion of some unknown kind of matters having large negative pressure so that strong energy condition ($\rho+3p>0$) is violated. Such an unknown matter is known as dark energy(DE).

In literature, there are various DE models to match with
observational data. The simplest model representing DE is the
Cosmological Constant which was introduced by Einstein himself,
surprisingly many years before the starting of DE craze. However,
this model of DE is not very popular due to many inherent
drawbacks (for example fine tuning problem ( Steinhardt 1997)). The other
candidates for DE are variable cosmological constant (Shapiro et. al. 2009; Sola et al. 2005 ; Solaet al. 2006
), the
canonical scalar field (Dutta et. al.  2009;Guo et al. 2007; Liddle
et. al. 1999;Ratra et al. 1988;Wetterich  1988;  Zlatev et al. 1999) (quintessence field), scalar field with
negative kinetic energy (phantom field) (Caldwell  2002 ; Caldwell et al. 2003; Nojiri et al.
2003B ;  Onemli et al. 2004; Saridakis  2009; Setare et al. 2008;
Setare et al. 2009) or a quintom field
(Capozziello et al. 2006; Elizalde et al. 2004; Feng et al.  2005;  Feng et al. 2006 ; Guo et al.  2005 ;  Li et al. 2005 ;  Setare 2006 ;   Setare et al. 2008A ;  Setare
et al. 2008B;  Setare et al. 2008C ;
Setare et al. 2009A; Zhao et al. 2006 ;) (a unified model of quintessence and phantom field). Further
a combined effort of quantum field theory and gravity leads to
speculate some nature of DE and is known
as holographic dark energy (HDE) model ( Copeland et al. 2006; Durrer et al. 2008;  Nojiri
et al., 2007; Padmanabhan 2002; Sahni   2005 ,2006; Nojiri et. al. 2006B).

In the present work we study the evolution of the horizons(apparent and event) for different DE models namely (a) DE with barotropic equationof state, (b) holographic DE(HDE) and (c) a non interacting two fluid system-HDE and dark matter in the form of dust. The paper is assigned as follows : Basic equations are presented in the section 2, evolution of the horizons are studied for the above three matter systems in section 3, section 4 deals with thermodynamical analysis of the universe bounded by the horizons. The paper ends withdiscussion and concluding remarks in section 5.

\section{Basic equations }

For simplicity let us start with homogeneous and isotropic model
of the universe (namely Friedmann-Robertson-Walker(FRW) model),
having line element

\begin{equation}\label{1}
ds^{2}=-dt^{2}+a^{2}(t)\left[\frac{dr^{2}}{1-kr^{2}}+r^{2}d\Omega^{2}\right]
\end{equation}

$$=h_{ab}dx^{a}dx^{b}+R^{2}d\Omega^{2}$$
where $$h_{ab}=diag\left(-1,
\frac{a^{2}}{1-kr^{2}}\right)~~~,~~~(a,~b=0,1~with~~x^{0}=t,
x^{1}=r)$$ and $$d\Omega^{2}=d\theta^{2}+sin^{2}\theta
d\phi^{2}~is~ the~ metric~ on~ unit~ two~ sphere.$$ $R=ar$ is the
radius of the sphere(area-radius), 'a' is the scale factor and
$k=0, \pm1$ stands for flat, closed and open model of our universe
respectively.

The matter is chosen as a perfect fluid with energy momentum
tensor
\begin{equation}\label{2}
T_{\mu\nu}=\left(\rho+p\right)u_{\mu}u_{\nu}-pg_{\mu\nu}
\end{equation}
So the Einstein field equations are (choosing $8\pi G=1=c$)
\begin{equation}\label{3}
H^{2}+\frac{k}{a^{2}}=\frac{1}{3}\rho
\end{equation}
\begin{equation}\label{4}
\dot{H}-\frac{k}{a^{2}}=-\frac{1}{2}\left(\rho+p\right)
\end{equation}
and the energy conservation equation is
\begin{equation}\label{5}
\dot{\rho}+3H\left(\rho+p\right)=0
\end{equation}

Combining (\ref{3}) and (\ref{4}) we get,
\begin{equation}\label{6}
\dot{H}+H^{2}=\frac{\ddot{a}}{a}=-\frac{1}{6}\left(\rho+3p\right)
\end{equation}

The dynamical apparent horizon which is essentially  the
marginally trapped surface with vanishing expansion, is defined as
a sphere of radius $R=R_{A}$ such that
\begin{equation}\label{7}
h^{ab}\partial_{a}R\partial_{b}R=0
\end{equation}
which on simplification gives
\begin{equation}\label{8}
R_{A}=\frac{1}{\sqrt{H^{2}+\frac{k}{a^{2}}}}
\end{equation}

The event horizon on the other hand is defined as (Davis 1998)
$$R_{E}=-a ~sinh(\tau)~~~~~~~~~~~~~~k=-1$$

\begin{equation}\label{9}
R_{E}=-a\tau~~~~~~~~~~~~~~~~~~~~~k=0
\end{equation}

$$R_{E}=-a~sin(\tau)~~~~~~~~~~~~~~~~~k=+1$$
where $\tau$ is the usual conformal time defined as

\begin{equation}\label{10}
\tau=-\int_{t}^{\infty}\frac{dt}{a(t)}~~~~~~~~~~~~~~~~~~~~~~~~~~~~~~|\tau|<\infty
\end{equation}

Note that if $|\tau |=\infty$, event horizon does not exist. Also
the Hubble horizon is given by
\begin{equation}\label{11}
R_{H}=\frac{1}{H}
\end{equation}
The horizons are related by the following relations (Mazumder
2009):
$$R_{A}=R_{H}<R_{E}~~~~~for K=0$$
$$R_{H}<R_{A}<R_{E}~~~~~~for K=-1$$
$$R_{A}<R_{E}<R_{H}~~~~~~~~~~~~~~~~$$
$$~~~~~~~~~~~~~~~or~~~~~~~~~~~~~~~~~for K=+1$$
$$R_{A}<R_{H}<R_{E}~~~~~~~~~~~~~~~~~$$

\section{Evolution of the horizons and consequences}

The time variation of the horizon radii are given by
\begin{equation}\label{12}
\dot{R}_{A}=-H\left(\dot{H}-\frac{k}{a^{2}}\right)R_{A}^{3}
\end{equation}
\begin{equation}\label{13}
\dot{R}_{E}=H R_{E}-\sqrt{1-\frac{k}{a^{2}}R_{E}^{2}}
\end{equation}
\begin{equation}\label{14}
\dot{R}_{H}=-\frac{\dot{H}}{H^{2}}
\end{equation}

One may note that the expression for $\dot{R_E}$ given in
references (Davis  1998) and (Mohseni Sadjadi  2006) are true only
for $k=0$. So the theorems given in the papers of Davis(1998) and
Sadjadi(2006) are only valid flat universe. However, in the
present work from the above expression (i.e., equation (13)) we
see that $R_{E}$ is an increasing or decreasing function of time
that depends only on whether $R_E>~or~<R_A$- it does not depend on
the nature of the matter involved as claimed by Davis and
Sadjadi.\\

We shall now study the variation of the horizons with the
evaluation of the universe. Due to observed accelerating phase of
the universe, the matter is assumed to be in the form of the DE
having equation of state $p=\omega \rho$.\\

{\bf Case I : $\omega$ is constant\\}
For simplicity, if we assume the flat model of the universe then from equation (\ref{6}) we have
$$\frac{\ddot{a}}{a}=\frac{\rho}{3}\left(1-\alpha\right)$$
with $\alpha=\frac{3}{2}\left(1+\omega\right)$. Hence in the quintessence era we have $0<\alpha<1$. Now solving the Einstein field equation (\ref{3}) and the conservation equation (\ref{5}) we have
\begin{equation}\label{15}
\alpha=a_{0}t^{\frac{1}{\alpha}},~~~\rho=\rho_{0}t^{-2}
\end{equation}
Then the horizons are given by
\begin{equation}\label{16}
R_{E}=\frac{\alpha t}{1-\alpha},~~~~~ R_{A}=\alpha t
\end{equation}
Hence, over one Hubble time $(t_{H}=\frac{1}{H})$ both have the same time variation, i.e.,
\begin{equation}\label{17}
t_{H}\frac{\dot{R_{h}}}{R_{h}}=\alpha~~~~~~(h\equiv E~or~A)
\end{equation}
Thus there are no significant changes of the two horizons over the Hubble time.\\

{\bf Case II : $\omega$ is variable\\} Here the choice of DE is
holographic model. The holographic principle states that the no.
of degrees of freedom for a system within a finite region should
be finite and is bounded roughly by the area of its boundary. From
the effective quantum field theory one obtains the Holographic
energy density as (Cohen et al. 1999)
\begin{equation}\label{18}
\rho_{D}=\frac{3c^{2}}{R_{E}}
\end{equation}
where free dimensionless parameter c is estimated from observation and IR cut off is chosen as $R_{E}$ to get correct expression.

Then using expression (\ref{18}) in the conservation equation (\ref{5}) the expression for the equation of state parameter is given by
\begin{equation}\label{19}
\omega=-\frac{1}{3}-\frac{2}{3}\sqrt{\frac{\Omega_{D}}{c^{2}}-\Omega_{k}}
\end{equation}
where $\Omega_{D}=\frac{\rho_{D}}{3H^{2}}$ and $\Omega_{k}=\frac{k }{a^{2}H^{2}}$ are the density parameters corresponding to DE and curvature respectively. Now from equations (\ref{12}) and (\ref{13}) the time variation of the horizons over one Hubble time are given by
\begin{equation}\label{20}
t_{H}\frac{\dot{R}_{A}}{R_{A}}=\frac{3}{2}\left(1+\omega\right)=1-\sqrt{\frac{\Omega_{D}}{c^{2}}-\Omega_{k}}
\end{equation}
and
\begin{equation}\label{21}
t_{H}\frac{\dot{R}_{E}}{R_{E}}=1-\sqrt{\frac{\Omega_{D}}{c^{2}}-\Omega_{k}}
\end{equation}
So both the horizons have the same time variation over one Hubble time.\\

{\bf Case III : Variable $\omega$ and two fluid syatem\\} Here we
consider a non-interacting two fluid system having one component
in the form of HDE and the other component as dark matter (in the
form of dust). So the Einstein equations for flat FRW model now
become

\begin{equation}\label{22}
H^{2}=\frac{1}{3}\left(\rho_{D}+\rho_{m}\right)
\end{equation}
\begin{equation}\label{23}
\dot{H}=-\frac{1}{2}\left(\rho_{D}+\rho_{m}+p_{D}\right)
\end{equation}
As the fluids are non-interacting so the energy conservation equations are
\begin{equation}\label{24}
\dot{\rho}_{m}+3H\rho_{m}=0
\end{equation}
and
\begin{equation}\label{25}
\dot{\rho}_{D}+3H\rho_{D}\left(1+\omega\right)=0
\end{equation}
So from the expression of the energy density for the HDE(given by equation (\ref{18})) we have as before
\begin{equation}\label{26}
\omega=-\frac{1}{3}-\frac{2}{3}\frac{\sqrt{\Omega_{D}}}{c}
\end{equation}
where variation of the density parameter is given by
\begin{equation}\label{27}
\Omega '=\Omega_{D}^{2}\left(1-\Omega_{D}\right)\left\{\frac{1}{\Omega_{D}}+\frac{2}{c\sqrt{\Omega_{D}}}\right\}
\end{equation}
Now the change of the horizons over one Hubble time are given by the expressions
\begin{equation}\label{28}
t_{H}\frac{\dot{R}_{A}}{R_{A}}=\frac{3}{2}-\frac{1}{2}\Omega_{D}-\frac{1}{c}\Omega_{D}^{\frac{3}{2}}
\end{equation}
\begin{equation}\label{29}
t_{H}\frac{\dot{R}_{E}}{R_{E}}=1-\frac{\sqrt{\Omega_{D}}}{c}
\end{equation}
Thus compared to between the the above two cases the changes of
the horizons over one Hubble time are not identical, though they
do not change significantly.\\

\section{Thermodynamics of the Universe and the role of the horizons : }
Here we consider the universe bounded by the event or apparent horizon as a thermodynamical system. In the previous section we have shown that neither the apparent horizon nor the event horizon change significantly over one Hubble time scale so equilibrium thermodynamics can be applied here with temperature and entropy on the horizon similar to black holes.\\
{\bf Case I : Matter in the form of perfect fluid :} \\
Here matter bounded by the horizon is considered to be in the perfect fluid. The total entropy change can be written as (for details see Mazumder et al 2009)
\begin{equation}\label{30}
\frac{d}{dt}\left(S_{I}+S_{h}\right)=\frac{4\pi R_{h}^{2}}{T_{h}}\left(\rho+p\right)\dot{R}_{h}
\end{equation}
where $R_{h}$ is the radius of the horizon(event or apparent),
$S_{I}$ and $S_{h}$ are respectively the entropy of the matter
bounded by the horizon and that of the horizon, $\rho$ and $p$ are
the energy density and the thermodynamic pressure of the inside
matter and $T_{h}$ is the temperature of the horizon as well as of
the inside matter for equilibrium thermodynamics. Thus generalised
second law of thermodynamics will be valid in quintessence era
$(\rho+p>0)$ if the radius of the horizon increases with time
while in phantom era $(\rho+p<0)$ the radius of the horizon should
decrease.\\

{\bf Case II : Matter in the form of HDE : }\\
If we differentiate the expression for the energy density of the HDE (i.e., equation (\ref{18})) then using the energy conservation equation (\ref{5}) we obtain (after simplification)
\begin{equation}\label{31}
\dot{R}_{E}=\frac{3}{2}HR_{E}\left(1+\omega\right)
\end{equation}
As before variation of the total entropy is given by equation (\ref{30}) which using (\ref{31}) becomes
\begin{equation}\label{32}
\frac{d}{dt}\left(S_{I}+S_{E}\right)=\frac{6\pi R_{E}^{3}}{T_{E}}\rho_{D}\left(1+\omega\right)^{2}
\end{equation}
for event horizon. For the apparent horizon using equation (\ref{12}) and the Friedmann equation (\ref{4}), equation (\ref{30}) simplifies to
\begin{equation}\label{33}
\frac{d}{dt}\left(S_{I}+S_{A}\right)=\frac{2\pi R_{A}^{5}H_{\rho}}{T_{A}}\rho_{D}\left(1+\omega\right)^{2}
\end{equation}
Thus generalised second law of thermodynamics hold for both the horizons when matter is purely in the form of HDE.\\

{\bf Case III : Non-interacting two fluid system : }\\
Here matter in the universe bounded by the horizon (event or apparent) is in the form of non-interacting two fluid system-one component is HDE $(\rho_{D}, ~p_{D})$ and the other is dark matter in the form of dust $(\rho_{m})$. Then total entropy variation (for details see Mazumder et. al. 2010) is given by
\begin{equation}\label{33}
\frac{d}{dt}\left(S_{I}+S_{A}\right)=\frac{4\pi R_{h}^{2}}{T_{h}}\left\{\rho_{m}+\rho_{D}\left(1+\omega\right)\right\}\dot{R}_{h}
\end{equation}
Thus energy density of dark matter plays a key role for the validity of the generalized second law of thermodynamics particularly in phantom era.

\section{Discussions and Concluding remarks:}

We shall now discuss the behavior of the horizons with the evolution of the universe both in
Quintessence and Phantom eras.
From the conservation equation (5) we see that in Quintessence era
$\rho$ is monotonic decreasing which reaches a local minima at the
phantom crossing and increases again with the evolution of the
universe as shown in Fig I. So the matter density has some short
of bouncing behavior at the phantom crossing. However, if the
universe starts contracting in phantom era (i.e., $H<0$) then
conservation equation demands $\rho$ should still decreases in the
phantom era and there is a point of inflexion at the phantom
barrier as shown in Fig.II . For both the possibilities in phantom
era $\rho$ has peculiar behavior when matter is exotic in nature
(i.e., $ \rho+p<0$). In the first case when universe is expanding
$\rho$ also increases in the phantom era indicating some matter
creation phenomena (of unknown nature) during that epoch. On the
other hand, when universe starts contraction in the phantom era,
$\rho$ still decreases, indicating destruction of mass in that
era.
\begin{figure}
\includegraphics[height=2.5in, width=3in]{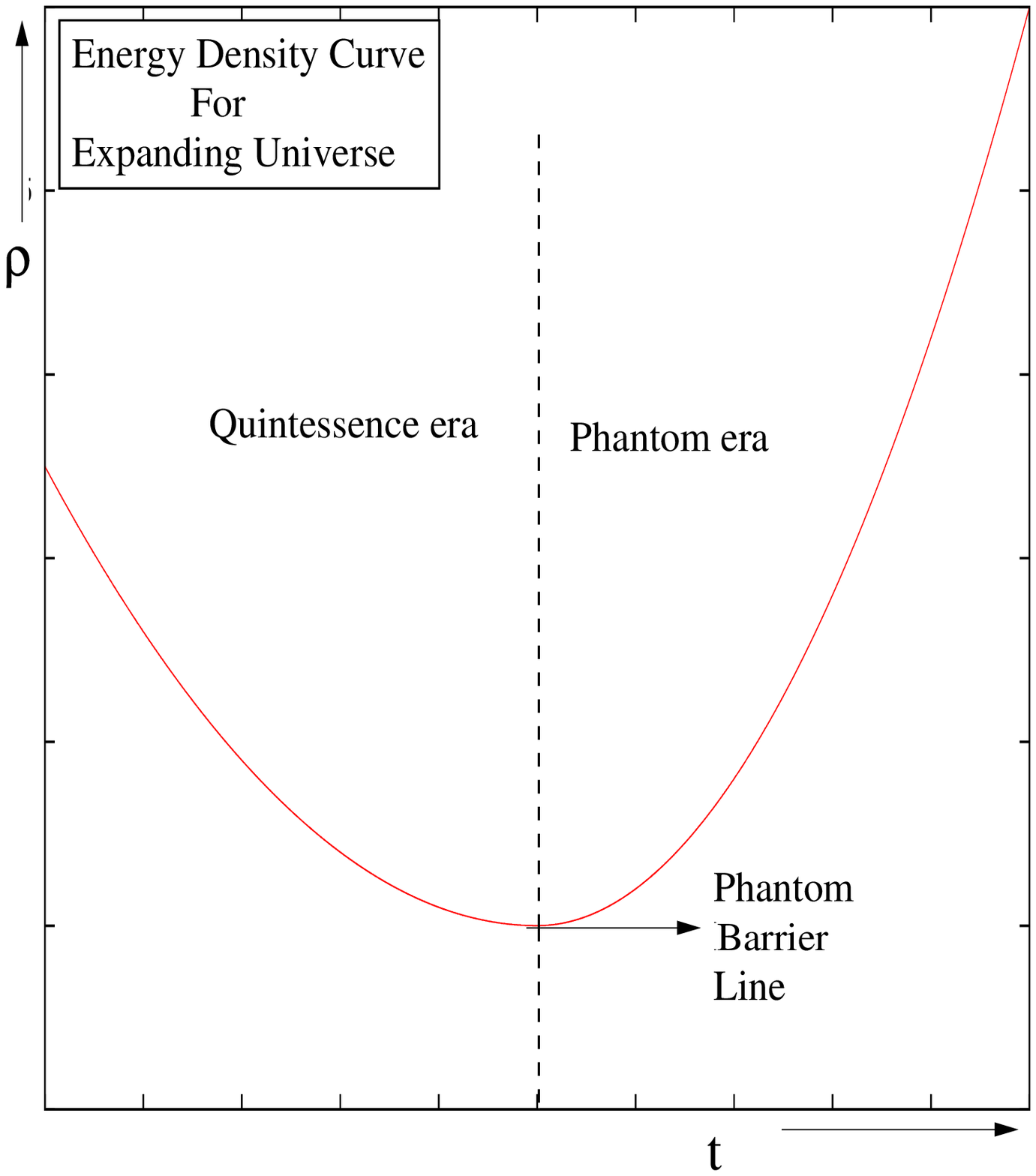}

~~~~~~~~~~~~~~~~~~~~~~~~~Fig.I  \hspace{1cm} \vspace{1mm}

Fig.I represents the variation of energy density with the
evolution of the universe in an expanding model. The dotted
vertical line denotes the phantom divide or phantom barrier line.
 \vspace{5mm}

\hspace{1cm}
\end{figure}

\begin{figure}
\includegraphics[height=2.5in, width=3in]{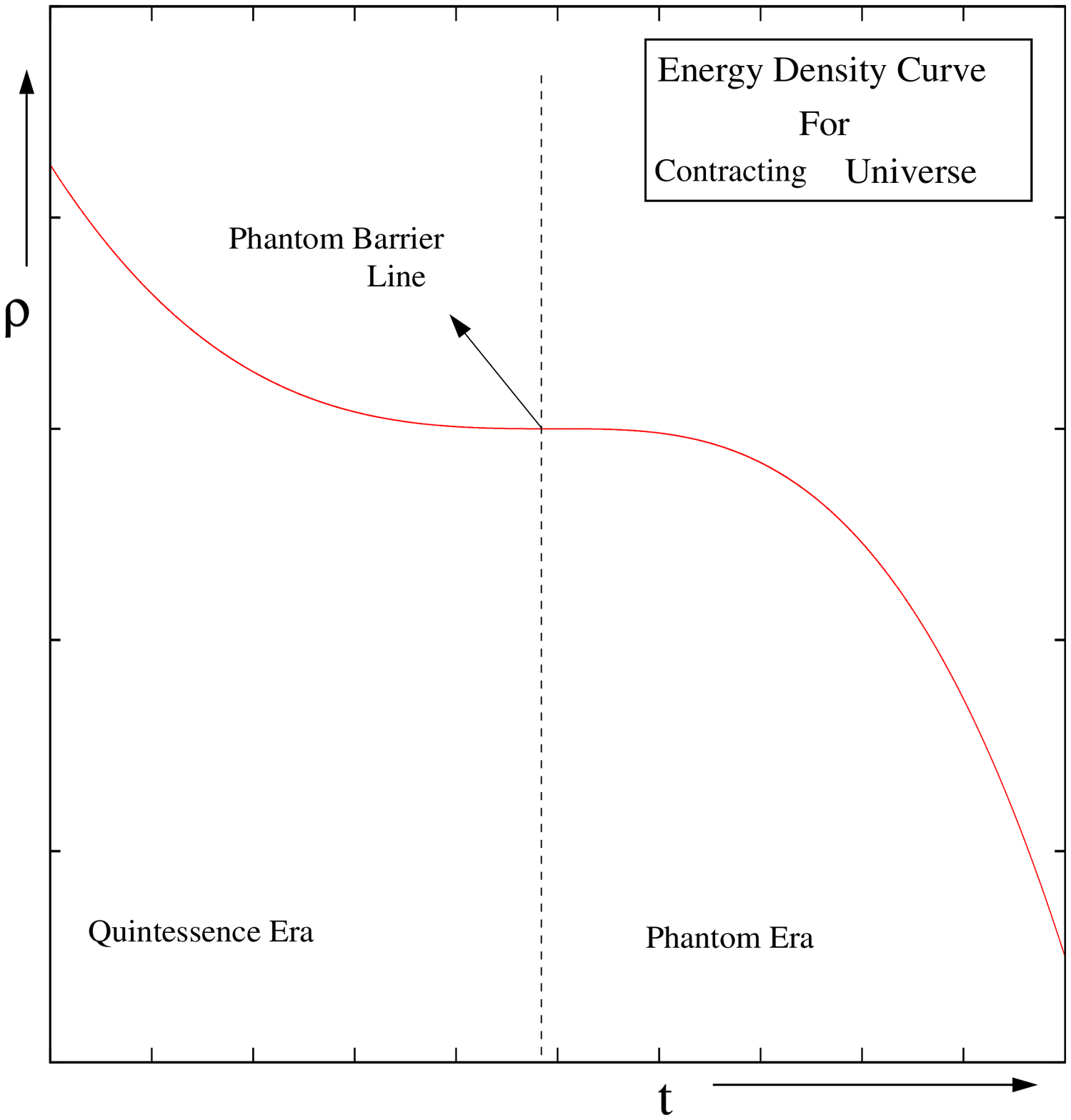}

~~~~~~~~~~~~~~~~~~~~~Fig.II \hspace{1cm} \vspace{1mm}

Fig.II represents the variation of energy density in an
contracting model of the universe in phantom era. \vspace{5mm}

\hspace{1cm}
\end{figure}
\begin{figure}
\includegraphics[height=2.5in, width=2.5in]{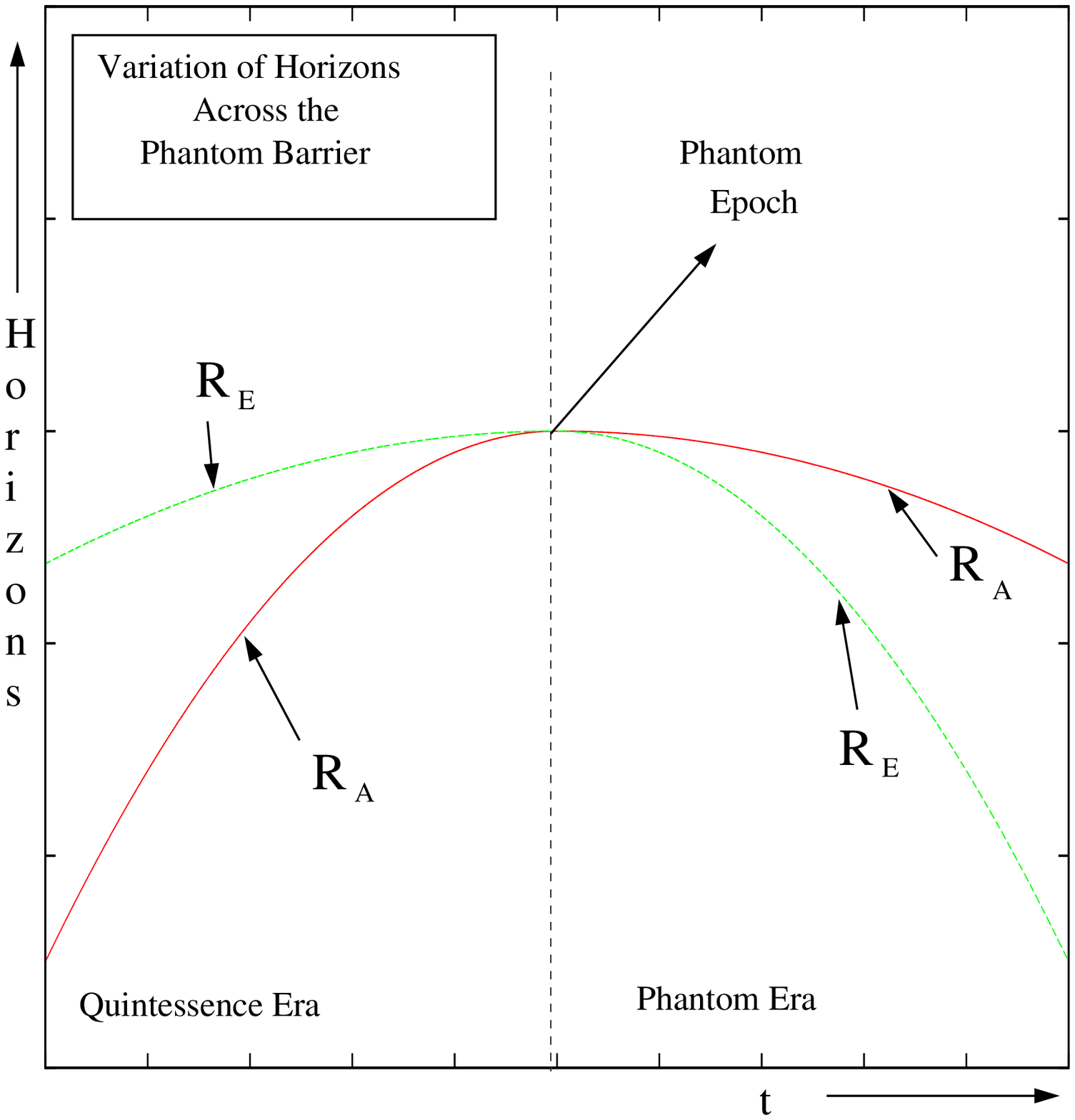}

~~~~~~~~~~~~~~~~~~~~~~~~Fig.III\hspace{1cm} \vspace{1mm}

Fig.III represents variation of event horizon and the apparent
horizon respectively in an expanding universe model. The dotted
vertical line again denotes the phantom divide line. As the
previous diagrams left side of which is denoting the quintessence
era whereas the right hand side represents the phantom era.
\vspace{5mm} \hspace{1cm}
\end{figure}

\begin{figure}
\includegraphics[height=2.5in, width=2.5in]{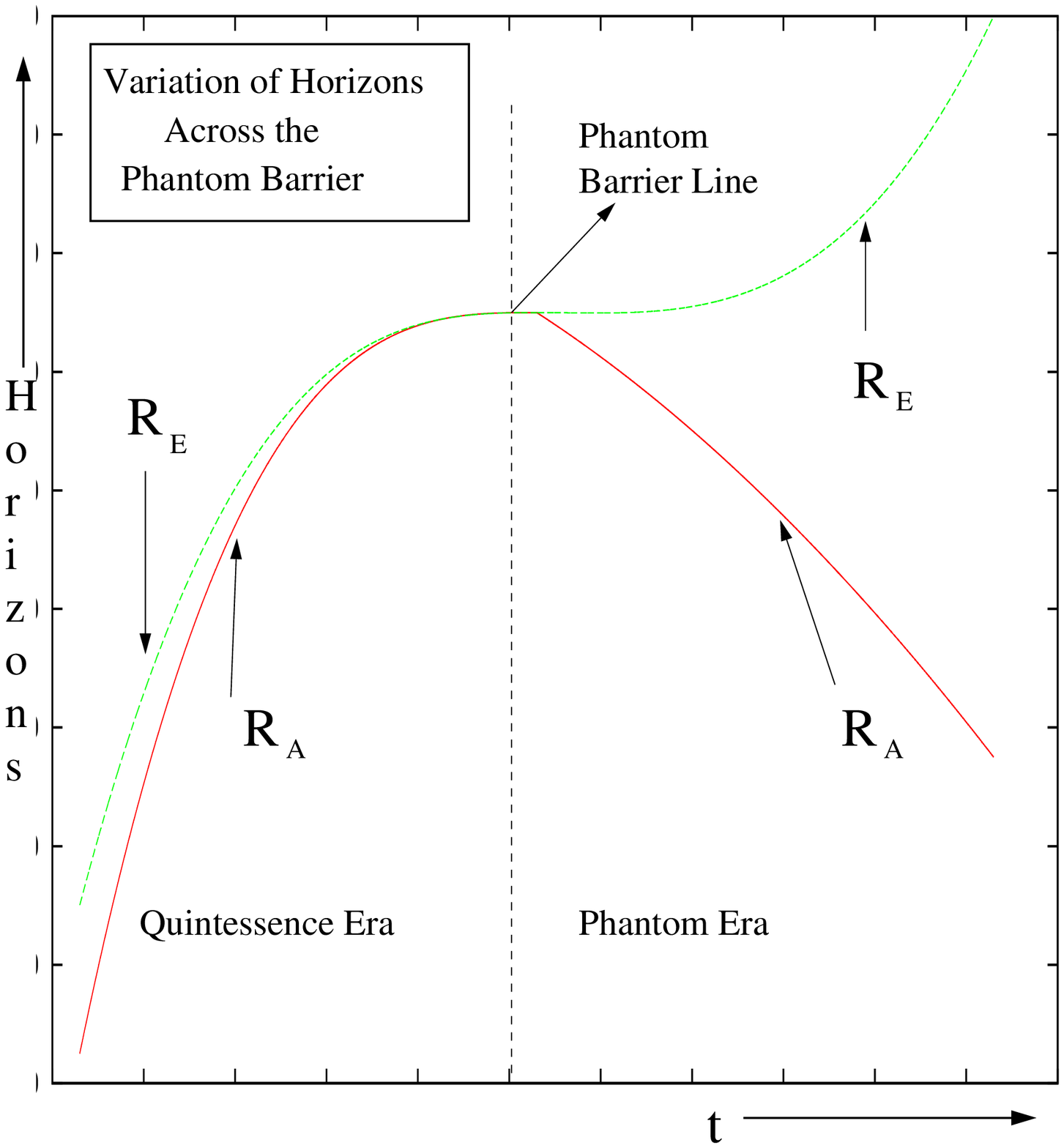}

~~~~~~~~~~~~~~~~~~~~~Fig.IV \hspace{1cm} \vspace{1mm}

Like Fig.III, Fig.IV also represents two curves showing the
variation of event horizon and apparent horizon. The dotted
vertical line denotes the phantom divide line. The left side of
which is denoting the quintessence era whereas the right hand side
represents the phantom era. \vspace{5mm} \hspace{1cm}
\end{figure}

We shall now present a comparative study of the evolution of the horizons across the phantom barrier with the expansion of the universe for the various DE matter distribution in the following tabular form\\

{\bf Table I:comparative study of the two horizons across phantom barrier}\\\\
\begin{tabular}{|c|c|c|}

\hline\hline  & Quintessence Era  & Phantom Era \\
\hline\hline Apparent  & (a)For HDE model apparent horizon  & (a) For HDE model $R_{A}$ decreases with time.
\\ Horizon & increases with the evolution of the universe. &
\\~~~~~ & (b) For other DE model the behavior & (b) $R_{A}$ has same behavior for other
\\~~~~~ &  is identical as HDE. & DE model as in HDE.\\
\hline~~ Event~~  & (a)For HDE model event horizon  & (a) For HDE
model $R_{E}$ decreases with time
\\ Horizon & increases and hence from equation(\ref{13}) & and so we have $R_{E}<R_{A}$. The variations
\\~~~~~ & we always have $R_{E}>R_{A}$. & are schematically shown in figure $III$
\\~~~~~ & (b) $R_{E}$ has similar behavior as in HDE, (for  & (b) For other DE model, $R_{E}$ may have
\\~~~~~ &  other DE models). & behavior as in HDE model. Also $R_{E}$  may
\\~~~~~ &  ~~ & still be an increasing function with the
\\~~~~~ &  ~~ &  evolution and the nature is schematically
\\~~~~~ &  ~~ & shown in figure $IV$.~~~~~~~~~~~~~~\\
\hline
\end{tabular}\\\\\\
We have also studied in the last section the thermodynamics of the
universe bounded by the horizons with different DE models. We have
assumed the validity of the first law of thermodynamics and
examined whether the GSLT  holds or not. The conclusions from this
thermodynamical study has been presented below in table II.\\

{\bf Table II: Validity of GSLT for different DE model}\\\\
\begin{tabular}{|c|c|c|}

\hline\hline  & Quintessence Era  & Phantom Era \\
\hline\hline Universe  & (a)For HDE model GSLT is always   & (a) GSLT is always satisfied for the HDE model.
\\ bounded & satisfied. &
\\by the & (b) For other DE model GSLT is satisfied & (b) Across the phantom barrier GSLT is satisfied.
\\apparent &  throughout the evolution. & (c) GSLT will be satisfied provided $\left|1+\omega\right|>\frac{\rho_{m}}{\rho_{D}}$
\\horizon &  (c) For non-interacting 2-fluid  system & ~~
\\ &  there is always validity of GSLT. &\\
\hline

 Universe~~  & (a)Thermodynamical system respects   & (a) No restriction is needed for the validity
\\ bounded & GSLT for HDE model. & of GSLT.
\\by & (b)GSLT is obeyed for other DE model. & (b) Validity of GSLT depends on the behavior
\\the  &  (c) No restriction is needed for the validity   &  of $R_{E}$. GSLT will be respected for the
\\event &  of GSLT for noninteracting 2-fluid system. &  fig $III$ while it will be violated for fig. $IV$.
\\horizon &  ~~ & (c) GSLT will be satisfied for variation
\\~~~~~ &  ~~ &  of $R_{E}$ according to fig $III$ provided
\\~~~~~ &  ~~ & $|1+\omega|~<~\frac{\rho_{m}}{\rho_{D}}$.\\
\hline

\end{tabular}\\\\\\

Thus, from the above study we see that both the evolution of the
horizons as well as the matter density have some strange behavior
in the phantom era, i.e., across the phantom barrier line.
therefore, for future work cosmological evolution in phantom
region will be done more in details and also it will be
interesting to explain the particle creation in the phantom era
with the expansion of the universe and possibly the mechanism of
particle creation may remove the possible future singularity.\\

\begin{contribution}

{\bf Acknowledgement :}\\

RB wants to thank West Bengal State Government for awarding JRF.
NM wants to thank CSIR, India for awarding JRF. All the authors
are thankful to IUCAA, Pune as this work has been done during a
visit.
\end{contribution}
\\\\

\frenchspacing {\bf REFERENCES} \frenchspacing
\\\\
Akbar M., Cai R.G. : {\it Phys. Lett B} {\bf 635} (2006) 7.\\
Allen, S. W. et al.: {\it Mon. Not. Roy. Astron. Soc.}, {\bf 353}, (2004), 457 .\\
A.G. Cohen , D.B. Kaplan and A.E. Nelson , {\it Phys. Rev. Lett.}
{\bf 82}
(1999) 4971.\\
Barrow J. D. : {\it Class. Quantum Grav.} {\bf 21} (2004)L79.\\
Bennett, C. L. et al.: {\it Astrophys. J. Suppl.} {\bf 148},(2003),1.\\
Cai, R. G.,  Kim , S. P. :{\it JHEP} {\bf 02} (2005) 050.\\
Caldwell, R. R. : {\it Phys. Lett. B} {\bf 545}, (2002) 23.\\
Caldwell, R.R.,  Kamionkowski, M., Weinberg, N. N. : {\it Phys. Rev.Lett.} {\bf 91}, (2003) 071301.\\
Capozziello, S. : {\it IJMPD} {\bf 11} (2002) 483.\\
Capozziello, S., Nojiri, S., Odintsov, S.D. :- {\it Phys. Lett. B} {\bf 632} 597 (2006)\\
Carroll, S. M.,  Duvvuri, V. ,Trodden, M. ,Turner, M.S. : {\it Phys. Rev. D} {\bf 68} (2004) 043528.\\
Copeland, E.J. , Sami, M., Tsujikawa , S. : {\it IJMPD } {\bf 15} (2006) 1753.\\
Davis, P.C.W. : {\it Class. Quantum Grav.} {\bf 5} (1998)1349.\\
Durrer, R. , Marteens , R. : {\it Gen. Rel. Grav.} {\bf 40} (2008) 301.\\
Dutta, S. ,Saridakis, E. N., Scherrer,  R. J. : {\it Phys. Rev. D} {\bf 79}, 103005 (2009).\\
Elizalde, E, Nojiri, S.,Odintsov, S. D. :-{\it Phys.Rev.D} {\bf 70} 043539(2004)\\
Feng, B. , Wang, X. L.,Zhang,  X. M. : {\it Phys. Lett. B}{\bf 607}, 35 (2005).\\
Feng, B., Li, M., Piao, Y.-S., Zhang, X. : {\it Phys. Lett. B} {\bf 634}, (2006) 101 .\\
Guo, Z. K. et al. : {\it Phys. Lett. B} {\bf 608}, (2005) 177.\\
Guo, Z. K.,Ohta, N., Zhang, Y. Z. : {\it Mod. Phys. Lett. A} {\it 22}, 883(2007).\\
Lancoz C. : {\it Ann. Math.} {\bf 39 } (1938) 842.\\
Li, M.-Z, Feng,  B.,Zhang,  X.-M : {\it JCAP}, {\bf 0512},(2005) 002.\\
Liddle, A. R., Scherrer, R. J. : {\it Phys. Rev. D} {\bf 59}, 023509 (1999)\\
Mazumder, N., Chakraborty, S. : {\it Class.Quant.Grav.} {\bf 26} 195016(2009).\\
Mazumder, N., Chakraborty, S. : {\it Gen.Rel.Grav.}{\bf 42} 813 (2010).\\
Nojiri, S.,  Odintsov, S.D. : {\it Phys. Rev. D} {\bf 68} (2003A)123512.\\
Nojiri, S. ,  Odintsov, S. D. : {\it Phys. Lett. B} {\bf 562},(2003B) 147.\\
Nojiri, S., Odintsov, S.D. : {\it Phys. Rev. D} {\bf 72}(2005) 023003.\\
Nojiri, S., Odintsov, S.D. : {\it Phys. Rev. D} {\bf 74}(2006A) 086005.\\
Nojiri , S., Odintsov , S. : {\it Gen.Rel.Grav.} {\bf 38} 1285,(2006B).\\
Nojiri S., Odintsov, S.D. :  {\it Int. J. Geom. Meth. Mod. Phys.} {\bf 4} (2007) 115.\\
Nojiri , S., Odintsov , S. : {\it arXiv: } {\bf 0801.4843A}[astro-ph].\\
Nojiri , S. , Odintsov , S. : {\it arXiv:} {\bf 0807.0685B} [hep-th].\\
Nojiri , S., Odintsov , S. : {\it arXiv: } {\bf 1011.0544v2 }[gr-qc].\\
Onemli, V. K. , Woodard, R. P. : {\it Phys. Rev. D} {\bf 70}, (2004) 107301.\\
Padmanabhan, T. : {\it Phys.Rept.} {\bf 380} (2002) 235.\\
Ratra B., Peebles, P. J. E. : {\it Phys. Rev. D} {\bf 37}, 3406(1988).\\
Riess, A. G. et al.: {\it AstroPhys J.} {\bf 607 } (2004)665.\\
Tegmark, M. et al.: [SDSS Collaboration], {\it Phys. Rev. D},{\bf 69}, (2004), 103501 .\\
Sadjadi, H.M. : {\it Phys. Rev. D} {\bf 73}(2006) 063525.\\
Sahni, V. : {\it AIP Conf. Proc.} {\bf 782} (2005) 166 .\\
Sahni, V. : {\it J. Phys. Conf. Ser.} {\bf 31} (2006) 115.\\
Saridakis, E. N. : {\it Nucl. Phys. B} {\bf 819}, (2009) 6116 .\\
Setare, M. R. : {\it Phys. Lett. B} {\bf 641},(2006) 130.\\
Setare, M. R., Sadeghi, J. , Amani, A. R. : {\it Phys. Lett. B} {\bf 666}, (2008A) 288.\\
Setare, M. R., Sadeghi, J., Amani,  A.R. : {\it Phys. Lett. B} {\bf 660}, (2008B) 299 .\\
Setare M. R., Saridakis, E. N. : {\it Phys. Lett. B} {\bf 668}, (2008C) 177.\\
Setare M. R., Saridakis, E. N. :  {\it JCAP} {\bf 0809},(2008D) 026 .\\
Setare M. R., Saridakis, E. N. :  {\it Int. J. Mod. Phys. D} {\bf 18},(2009A) 549 .\\
Setare, M. R.,Saridakis, E. N. : {\it JCAP} {\bf 0903},(2009B)002.\\
Shapiro, I. L.,  Sola, J. : {\it Phys. Lett. B} {\bf 682}, (2009) 105 .\\
Sola, J., Stefancic, H. : {\it Phys. Lett. B} {\bf 624},(2005) 147.\\
Sola, J.,Stefancic,  H. : {\it Mod. Phys.Lett.A} {\bf 21}, (2006) 479.\\
Steinhardt, P. J.{\it Critical Problems in Physics} (1997),Princeton University Press.\\
Wetterich,C. : {\it Nucl. Phys. B} {\bf 302}, 668 (1988).\\
Zhao, W., Zhang, Y. : {\it Phys.Rev. D} {\bf 73},(2006) 123509.\\
Zlatev, I.,Wang, L. M., Steinhardt, P. J. : {\it Phys. Rev. Lett.} {\bf 82}, 896 (1999).\\

\end{document}